%% file: Meyers_Soiling.tex
\documentclass[conference]{IEEEtran}
\IEEEoverridecommandlockouts
\input defs.tex
\usepackage{cite}
\usepackage{amsmath,amssymb,amsfonts,siunitx}
\usepackage{algorithmic}
\usepackage{graphicx}
\usepackage{textcomp}
\usepackage{xcolor}
\usepackage{pgf,pgfplots}
\usepackage{import}
\usepackage{booktabs}
\def\BibTeX{{\rm B\kern-.05em{\sc i\kern-.025em b}\kern-.08em
    T\kern-.1667em\lower.7ex\hbox{E}\kern-.125emX}}
\begin{document}

\title{Estimation of Soiling Losses in Unlabeled PV Data
\thanks{This material is based on work supported by the U.S. Department of 
Energy's Office of Energy Efficiency and Renewable Energy (EERE) under the 
Solar Energy Technologies Award Number 38529.}
}

\author{\IEEEauthorblockN{Bennet Meyers$^{1,2}$}
	\IEEEauthorblockA{$^1$ SLAC National Accelerator Laboratory, Menlo Park, 
	CA, 94025, USA \\$^2$ Stanford University, Stanford, CA, 94305, USA}}

\maketitle

\begin{abstract}
We provide a methodology for estimating the losses due to soiling for 
photovoltaic (PV) systems. We focus this work on estimating the losses from 
historical power production data that are unlabeled, \ie~power 
measurements with time stamps, but no other information such as site 
configuration or meteorological data. We present a validation of this approach 
on a small fleet of typical rooftop PV systems. The proposed method differs 
from prior 
work in that the construction of a performance index is not required to analyze 
soiling loss. This approach is appropriate for 
analyzing the soiling losses in field production data from fleets of 
distributed rooftop systems and 
is highly automatic, allowing for scaling to large fleets of heterogeneous PV 
systems. 
\end{abstract}


\section{Introduction}\label{s-intro}


Soiling can cause significant energy yield reduction in photovoltaic (PV) 
systems, as high as 
$-1\%$/day, but 
the effects are quite variable and impacted by many factors from system 
geometry to local climate conditions to nearby industry and 
agriculture~\cite{Ilse2019}. Quantifying the losses due to soiling is important 
for understanding and mitigating this effect, which in turn improves the 
overall reliability and dependability of PV as an energy generation source. 
This importance is seen in the dedicated subarea for soiling at this 
conference, as well as the strong emphasis on soiling at other events such as 
the yearly PV Reliability Workshop hosted by NREL. Largely absent from this 
conversation, however, is quantification of the impacts of soiling in large 
fleets of heterogenous, distributed PV systems, which comprised over 40\% of 
the installed capacity in 2020~\cite{SEIA2021}. The reason for this is a 
technical one: it is very difficult to analyze soiling trends without a 
reliable reference, and the unlabeled nature of distributed PV data make 
generating this reference difficult or impossible.

In this work, we present a methodology, based on recent research on machine 
learning for signal 
processing, that extracts estimates of system soiling losses from 
\emph{unlabeled} production data. This 
work eliminates the requirement of constructing a performance index to analyze 
soiling loss and is highly 
automated, thus enabling the large-scale analysis of fleets of thousands of 
heterogenous systems. The 
proposed approach is based on an implementation of the \emph{signal 
decomposition} 
(SD) framework~\cite{Meyers2022}.

Our approach takes  unlabeled PV power generation 
measurements as an input and returns an estimate 
of the soiling loss over time, given as a percent loss relative to the unsoiled 
performance. This trend may be used to calculate secondary statistics such as 
the total energy loss or seasonal loss patterns. We validate this method on 
synthetic data, labeled data from a soiling test site, and on representative 
unlabled data. The algorithm is available as a module in the Solar Data Tools 
package~\cite{Meyers2020b, solar-data-tools}. This approach is uniquely suited 
to the analysis of fleet-scale PV systems, where it can be difficult or 
impossible to get suitable reference data for normalization.

\section{Related work and contributions}

We are not the first to propose a method for estimating soiling losses from PV 
production data~\cite{deceglie2018,skomedal2019,Skomedal2020,Micheli2021}. As 
described 
in~\cite{Skomedal2020}, it has been determined that a combined model of 
degradation, soiling, and seasonal bias outperforms estimates of each loss term 
separately.

\begin{figure}
\centering
\resizebox{\columnwidth}{!}{
\import{figs/}{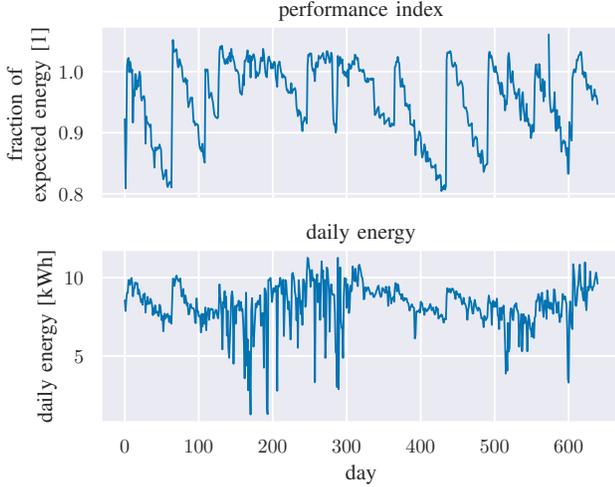}
}
\caption{A comparison of labeled data (top) and unlabeled data (bottom) from a 
single PV system in a desert environment.}
\label{f-labeled-unlabeled}
\end{figure}

In this paper, we build upon previous work in a few ways. First, we propose a 
clear unified signal model (implemented as an SD problem) to describe the 
underlying 
components, rather than invoking iterative heuristics. Second, the use of the 
extensible SD framework allows for the expression of component classes that are 
designed specifically for this application  Third, our approach 
estimates soiling losses from unlabeled system power generation time series 
data, without the need to construct a performance index (PI). As discussed in 
the introduction, this unlocks the analysis of around 40\% of the installed PV 
capacity in the United States. An example of a 
PI constructed from data labels and an unlabeled daily energy signal are shown 
in figure~\ref{f-labeled-unlabeled}. The proposed method allows for the 
analysis of \emph{both} unlabeled and labeled data. Labeled data will provide 
more accurate results, when available, but reasonable estimates may be obtained 
when such labels are unavailable. 

\section{Methods}

We construct an SD problem~\cite{Meyers2022} that models the decomposition of 
measured daily PV system energy or normalized energy (\ie, 
a performance index, ``PI'') 
into a number of components, one of which 
represents the soiling loss in the system. This approach can be thought of as 
an \emph{unsupervised} machine learning (ML) method for finding structure in 
time series data, similar to model-based clustering methods like Gaussian 
mixture models~\cite[\S14.3.7]{Hastie2009}. Unlike supervised ML, there is no 
``training'' of the method; we simply design the mathematical optimization 
problem, input the data for analysis, and receive the soiling estimate. In this 
section, we describe the 
data preparation, SD formulation, and validation procedure. Finally, we 
briefly describe a procedure for utilizing the estimate of system soiling to 
``correct'' the measured power data.

\subsection{Data preparation}

The proposed method of signal decomposition operates on a discrete daily time 
series representing raw or normalized daily system energy production, which we 
refer to as ``unlabeled'' and ``labeled'' data respectively. The purpose of 
generating a performance index is to remove known sources or variation in data, 
particularly due to available irradiance and operating temperature, and is a 
typical analytical approach for PV performance analysis, but requires 
additional knowledge about the system and its operating environment beyond real 
power production. 

The raw data may be any measurements of PV system power 
or energy production indexed in time. This may be 1-minute measurements of 
instantaneous power, 15-minute interval averaged power, or daily energy 
production. If starting from high-frequency power 
measurements, one simply integrates to get the daily energy production. If one 
wishes to construct a PI, you then normalize by expected daily energy at this 
time.

When starting from sub-daily measurements, any prefiltering step may be 
applied, and rejected days can have their values replaced with \texttt{NaN} 
values (which we represent as $?$ in our notation). The SD framework optimally 
handles missing data points, so there is no 
need to use corrupted or untrustworthy data nor replace such data with 
interpolated values.

In this paper, we validate on both synthetic and real data. As described 
in~\S\ref{s-synthetic-data}, 
the synthetic data is already a daily time-series (a normalized PI), so no 
preprocessing is 
required. For the real data, which has a 5-minute measurement interval, we use 
the data cleaning and filtering tools provided in Solar Data 
Tools~\cite{Meyers2020b, solar-data-tools}, and replace days that do not pass 
the quality check 
with $?$ values. After obtaining a representation of daily energy 
production (possibly with missing values), the signal is scaled so the 
95$^\text{th}$ percentile is equal to 1. This is our input to the 
\textit{signal decomposition problem} (SD problem). We do not construct a 
performance index on the real data and instead analyze the raw energy signal.

\subsection{SD problem formulation}

Utilizing the notation of signal decomposition defined in~\cite{Meyers2022}, we 
say that our data for the SD problem is a signal $y\in\left(\reals 
\cup \{?\} \right)^{T\times p}$ with length 
$T$ equal to the number of days in the data set and measurement dimension 
$p=1$. In this case, because the 
measurement dimension is equal to 1, $y$ may also be thought of as a column 
vector of 
length $T$. We model the signal $y$ as the 
composition of $K=4$ components, $x^1$ to $x^4$, the sum of which must be equal 
to the signal $y$ at the entries that do not contain missing values, or in 
other words,
\[y_t  = x_t^1 + x_t^2 + x_t^3 + x_t^4,\mbox{ for }t\in\mathcal{K},\]
where $\mathcal{K}$ is the set of time indices that do not contain missing 
values (the ``known'' set). The four 
components are defined in the SD model by their cost functions $\phi_k(x^k)$ 
for 
$k=1,\ldots,4$. (We drop the superscript $k$ on $x$ to keep the notation 
lighter when not distinguishing between particular components.) As we will see, 
the last component, $x^4$, will represent the soiling signal which we wish to 
estimate from the data. The other three components represent other processes 
which impact the 
energy production of the system.

\paragraph{Component definitions}
The first component represents 
the residual of the model, and it is taken to be the quantile cost 
function~\cite{Koenker1978,Koenker2001},
\[\phi_1(x) = \mathbf{quant}_\tau(x) = \sum_{t=1}^T (1/2)\left\lvert x_t 
\right\rvert + (\tau - 1/2) 
x_t,\]
where $\tau\in(0,1)$ is a parameter. When $\tau<0.5$, positive residuals are 
prefered to negative residuals, and vice versa when $\tau>0.5$. When working 
with raw energy data, we set $\tau=0.85$, which strongly prefers negative 
residuals, accounting for the fact that we have not normalized for weather 
effects, and clouds tend to reduce rather than increase the system energy 
product. When operating on normalized, PI data (such as the synthetic data set 
in this paper), we set $\tau=0.5$ since we expect the deviations from the 
expected output to be symmetric, or at the very least more symmetric then when 
no normalization is performed.

The second component is 
a seasonal term, which is smooth and periodic each year,
\[\phi_2(x) =  \left\{ \begin{array}{ll}
\lambda_2\| D_2 x\|_2^2 & x_t = x_{t+Y}\mbox{, for }t=1,\ldots T-Y\\
\infty & \mbox{otherwise,}
\end{array}\right.\]
where $D_2\in\reals^{(T-2)\times T}$ is the second-order discrete difference 
operator. (See, for example, \cite[\S6.4]{Boyd2018} for information on 
difference matrices.) $\lambda_2$ is a weighting 
parameter, and $Y=365$ is the period of 
the component. The normalization of the energy signal affects the expected 
amplitude of this component. That is, we would expect raw energy data to have a 
larger seasonal component than normalized data. The component definition 
presented here covers both cases well and is not sensitive to the amplitude of 
the component.

The third component represents the bulk, long-term degradation 
rate, and is given by
\[\phi_3(x) =  \left\{ \begin{array}{ll}
0 & x_0 = 0\mbox{ and }x_t = mt + b\mbox{ for }t=1,\ldots,T\\
\infty & \mbox{otherwise,}
\end{array}\right.\]
for some values of $m$ and $b$. This just constraints the component to be 
linear with respect to time with an initial value equal to 
zero. We have chosen a linear degradation model due to its popularity in the 
literature, but we note that other trend models could be employed within the SD 
framework, \eg, a smooth, monotonically decreasing signal. Note that the third 
component does not include a weight parameter, as the 
penalty function only takes on values of zero or infinity. 

The fourth and final 
component represents what we are interested in measuring, the soiling losses 
in the system. This cost is defined as,
\[\phi_4(x) =\left\{ \begin{array}{ll}
\ell_{4a}(x) + \ell_{4b}(x) + \ell_{4c}(x) & x \preceq 0\\
\infty & \mbox{otherwise,}
\end{array}\right.\]
where the summed functions are
\BEAS
\ell_{4a} &=& \lambda_{4a}\|D_2 x \|_1 \\
\ell_{4b} &=& \lambda_{4b}\sum_{t=0}^T(-x_t) \\
\ell_{4c} &=& \lambda_{4c} \mathbf{quant}_\tau(D_1 x)
\EEAS
with parameters $\lambda_{4a}$, $\lambda_{4b}$, and $\lambda_{4c}$. $D_2$ 
is again the 
second-order 
discrete difference operator, and $D_1$, similarly, is the first-order 
difference. The quantile cost parameter $\tau$ is taken to be $0.9$ here. This 
cost is a composite of simpler functions, which combine to 
select for signals with the following characteristics:
\begin{itemize}
\item non-positive (soiling can only reduce the system power)
\item sparse in second-differences (\ie, piecewise linear)
\item with values ``close'' to zero (is sum-absolute sense)
\item a preference for more values with a negative slope than a positive one 
(\ie, a tendency towards slow degradation and quick recovery).
\end{itemize} 

Component cost $\phi_4$ demonstrates the \emph{extensibility} of the SD 
framework. We are able to build up a complex cost function from smaller units 
and design it in a way to capture domain knowledge about the component. 

\paragraph{SD parameters}
The SD problem formulating includes four parameters, $\lambda_2$, 
$\lambda_{4a}$, $\lambda_{4b}$, and $\lambda_{4c}$. These parameters are 
tunable, and different values can greatly effect the characteristics and 
quality of the resulting decomposition. A deep discussion on the role of 
parameters in SD problems can be found in~\cite[\S2.6]{Meyers2022}. While a 
method is provided in~\cite[\S2.7]{Meyers2022} for selecting optimal parameter 
values, we find that in this context it makes more sense to rely on the 
practical experience of the analyst. In other words, we have found values 
that work well in many cases, and a small amount of hand-tuning is accepted in 
other cases. A description of these parameters and their default values are 
given in table~\ref{t-params}.

\begin{table}
\centering
\caption{SD problem parameters}
\begin{tabular}{ccl}
\toprule
param. & value & description \\\midrule
$\lambda_2$ & \num{5e2} & stiffness of seasonal baseline  \\
$\lambda_{4a}$ & \num{2} & effects number of soiling component breakpoints  \\
$\lambda_{4b}$ & \num{3e-2} & penalizes large values of soiling component  \\
$\lambda_{4c}$ & \num{2e-1} & encourages asymmetric rates in soiling component  
\\\bottomrule
\end{tabular}
\label{t-params}
\end{table}

\paragraph{Solution method}
The SD problem is convex (inequality-constrained quadratic 
program~\cite[\S4.4]{convex_opt}) and of modest size (around 2.1k variables for 
3-year 
data set to around 15k variables for a 10-year data set), so we 
simply use CVXPY~\cite{diamond2016cvxpy,agrawal2018rewriting} and the 
commercial Mosek solver~\cite{Andersen2000}, which is sufficient for research 
purposes. An implementation of the algorithm described in~\cite{Meyers2022}
would allow for the removal of the dependence on Mosek and is an area for 
future work. 

\subsection{Validation}\label{s-validation}


The most desirable method of validation for an unsupervised machine learning 
algorithm such as the methods described in this paper would be access to real 
PV system production data that has been hand-labeled with soiling trends. 
Because this is difficult to obtain or generate, we take a multi-modal approach 
to validation in this paper, with three different approaches to validation, 
described below, ordered by how well labeled the data source is. 

\paragraph{Synthetic data}\label{s-synthetic-data}
We follow the methods published in~\cite{Skomedal2020} to evaluate the 
performance of the algorithm on synthetic data that represents normalized 
energy productions. This approach generates random realizations of PI signals, 
with components drawn from pre-defined statistical models. The synthetic data 
model includes a `seasonal' term which 
represents the seasonal variation in performance. Normalization with a 
performance index is typically expected to lower this seasonal variation, but 
it does not fully remove it. The 
noise term in the model is Gaussian white noise, 
representing the assumption that a PI signal normalized for weather phenomenon. 
We therefore set $\tau=0.5$ in $\phi_1$ of the SD formulation to reflect the 
expectation of symmetric residuals.

Because the synthetic model explicitly models the system soiling losses, we are 
able to directly assess the ability of the SD soiling algorithm to estimate the 
hidden soiling signal. We select \emph{mean-absolute error} (MAE)
as a summary error metric for comparing the known synthetic soiling loss to the 
SD estimate, which is preferred over \emph{root-mean-square error} when the 
errors are not expected to be normally distributed~\cite[\S6.1.2]{convexopt}. 
Because analysts often want to be able to estimate soiling \emph{rates} on a PV 
system, in addition to understanding the total energy loss, we calculate the 
MAE on both the soiling loss component and the first-order 
difference~\cite{numpy-diff} of the loss, which we call the soiling rate. 
Finally, we note a small error in estimate of a cleaning event (\ie, one day 
before or after the true event) is of small consequence to the analyst but will 
result in very large error values, especially for the analysis of soiling 
rates. Therefore, we introduce a third summary statistic which is the MAE of 
\emph{filtered} soiling rate, which simply selects for time periods when both 
the synthetic soiling component and the SD estimate agree that the instantanous 
soiling rate is negative, \ie, neither component is currently in a cleaning 
event. This final metric provides useful insight into the ability of the 
algorithm to accurately estimate the rate of soiling loss between cleaning 
events.


\paragraph{Labeled production data}
Labeled soiling data was presented in~\cite{Kam-Lum2021} and analyzed for 
soiling trends. We leverage these published results to validate the SD soiling 
algorithm on the \emph{unlabeled} power production data. For this power, we 
ignore the reference system, treating the 
test system as an unlabeled data source, and estimate the soiling 
losses of the test system using the SD formulation. Then we compare results of 
the proposed method to the results from~\cite{Kam-Lum2021}. We calculate the 
same three error metrics for this data set, as described previously for the 
synthetic data.

\paragraph{Unlabeled production data}
We demonstrate the application of the SD soiling method on a selection of 50 PV 
systems, with only access to measured real power. There is no known soiling 
trend to compare to in this case. Instead, this represents what we see as a 
typical use case, and we show how outlier sites may be identified.

\subsection{Soiling correction for downstream analysis}
We briefly note that the estimate of soiling loss may be used to ``correct'' 
for soiling for other analysis. For example, we note in two other papers 
submitted to this conference~\cite{pvsc-shade,pvsc-shade-david}, that it is 
beneficial to account for soiling in production PV data prior to analyzing 
shade losses. We provide a simple procedure for doing this correction here.

The soiling loss component is a daily signal representing the fractional loss 
in daily energy from soiling, typically between 0 and 1. Therefore, the 
expected power output of the system in the absence of soiling would be the 
measured power divided by the instantaneous soiling loss. The Solar Data Tools 
implementation of the methods described in this paper includes a feature for 
automatically performing this correction.

\section{Results}

\subsection{Synthetic data}\label{s-results-synthetic}

\begin{figure}
\centering
\resizebox{\columnwidth}{!}{
\import{figs/}{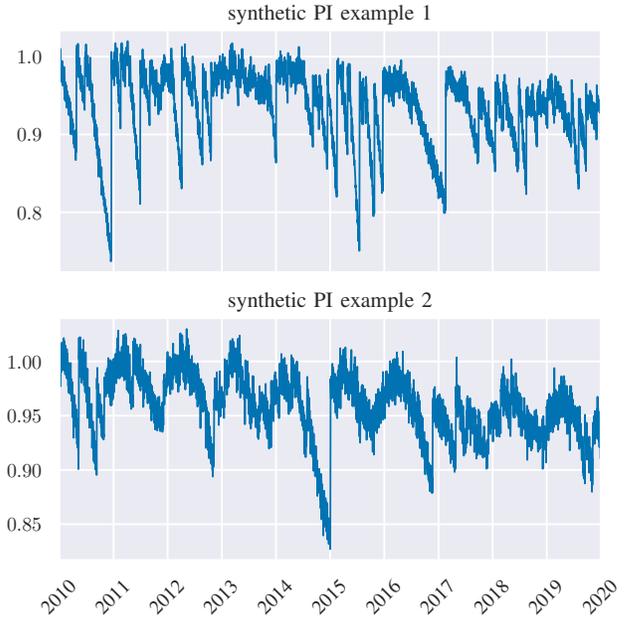}
}
\caption{Two typical synthetic soiling signals generated by NREL software. The 
top signal is drawn from scenario 1 and the bottom from scenario 2.}
\label{f-synthetic-soil}
\end{figure}

\begin{figure}
\centering
\resizebox{\columnwidth}{!}{
\import{figs/}{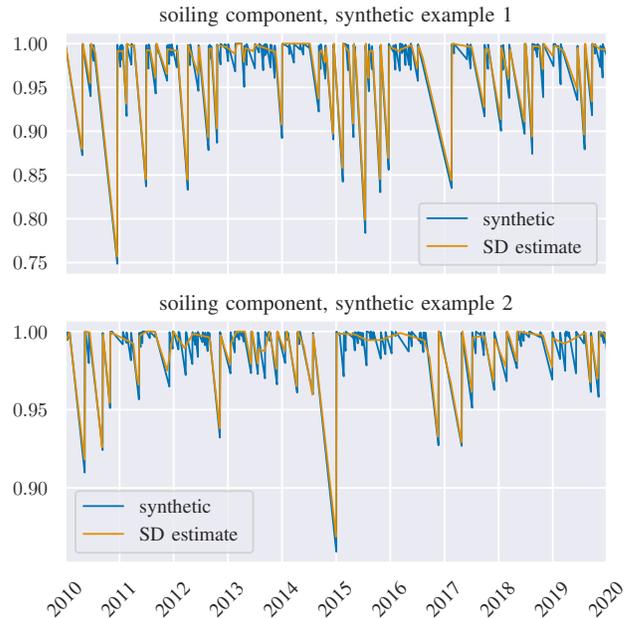}
}
\caption{Comparison of the actual and estimated soiling components in the two 
synthetic examples shown in figure~\ref{f-synthetic-soil}.}
\label{f-synthetic-results}
\end{figure}

Following the scenario generation procedure defined in~\cite[\S 
II-D]{Skomedal2020}, we define 
six generative model configurations, with different levels of soiling, 
seasonality, and noise. These scenarios are briefly described 
in~\ref{t-scenarios}. Two characteristic examples of the synthetic PI signals 
are shown in figure~\ref{f-synthetic-soil}, and the SD estimates of the soiling 
trends are compared to the true values in figure~\ref{f-synthetic-results}. The 
error metric for these two examples is given in table~\ref{t-synth-case}. The 
soiling rate were on average around 
-0.001 for the first example and -0.0005 for the second example.

\begin{table}
\centering
\caption{Synthetic data scenarios}
\begin{tabular}{c*{2}{l}}
\toprule
number& name & description \\\midrule
1 & normal & soiling rates ($sr$) uniform in $[0, 0.003]$ \\
2 & M soil, H season & $sr\in[0, 0.001]$, double seasonal 
amplitude \\
3 & M Soil, H noise & $sr\in[0, 0.001]$, double noise 
amplitude \\
4 & seasonal cleaning & $sr\in[0, 0.005]$, cleaned seasonally \\
5 & M soil & $sr\in[0, 0.005]$ \\
6 & L soil & $sr\in[0, 0.001]$ \\\bottomrule
\end{tabular}
\label{t-scenarios}
\end{table}
\begin{table}
\centering
\caption{Error metrics for two synthetic soiling examples}
\begin{tabular}{l*{4}{c}}
\toprule
&loss MAE & rate MAE & filtered rate MAE \\\midrule
Ex. 1 & 0.008698 & 0.002257 & 0.000379 \\
Ex. 2 & 0.005366 & 0.000919 & 0.000202 \\\bottomrule
\end{tabular}
\label{t-synth-case}
\end{table}

\begin{figure}
\centering
\resizebox{\columnwidth}{!}{
\import{figs/}{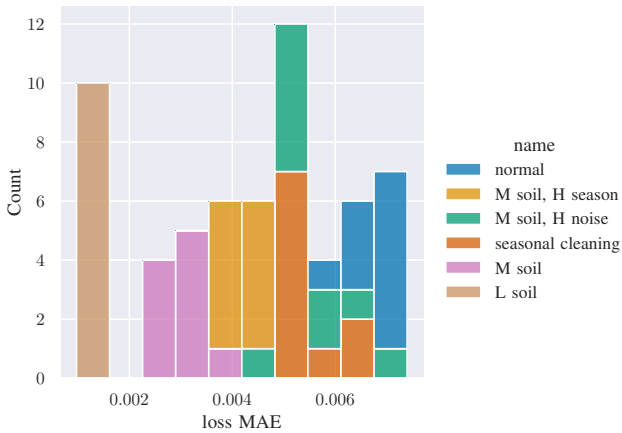}
}
\caption{Distribution of loss MAE for the 60 realizations, labeled by scenario.}
\label{f-synth-loss-mae}
\end{figure}

\begin{figure}
\centering
\resizebox{\columnwidth}{!}{
\import{figs/}{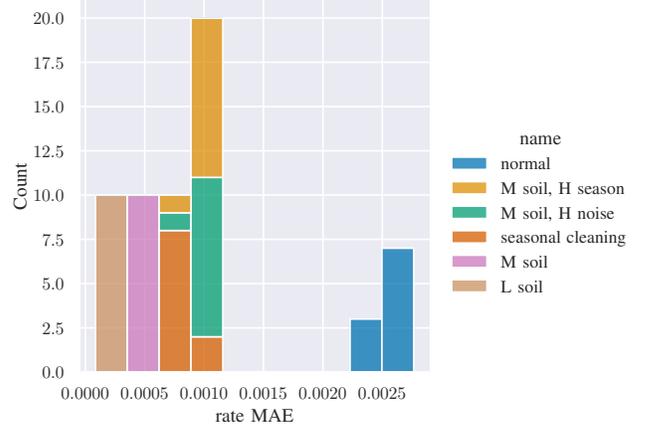}
}
\caption{Distribution of rate MAE for the 60 realizations, labeled by scenario.}
\label{f-synth-rate-mae}
\end{figure}

\begin{figure}
\centering
\resizebox{\columnwidth}{!}{
\import{figs/}{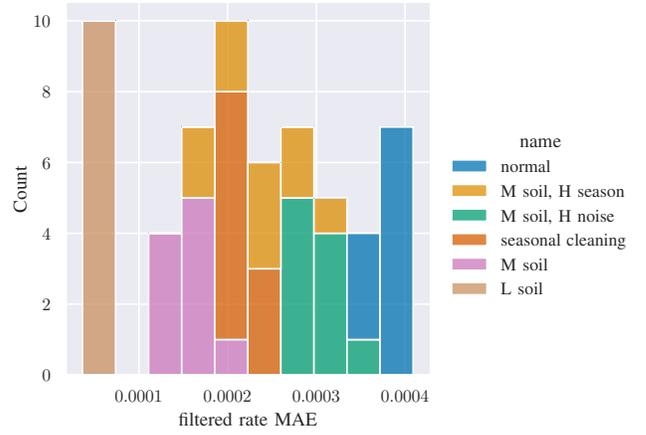}
}
\caption{Distribution of the filtered rate MAE for the 60 realizations, labeled 
by scenario. This metric accurately captures how closely the SD method 
estimated the loss rates during soiling periods.}
\label{f-synth-frate-mae}
\end{figure}

We sample the 6 scenarios 10 times each, and we
solve the associated SD problem for each of the 60 realizations. Finally, we 
calculate the three error metrics for each realization. The distribution of 
loss MAE is given in figure~\ref{f-synth-loss-mae}; the distribution of rate 
MAE is given in figure~\ref{f-synth-rate-mae}, and the distribution of the 
filtered rate MAE is given in figure~\ref{f-synth-frate-mae}.


\subsection{Labeled data}
\begin{figure}
\centering
\resizebox{\columnwidth}{!}{
\import{figs/}{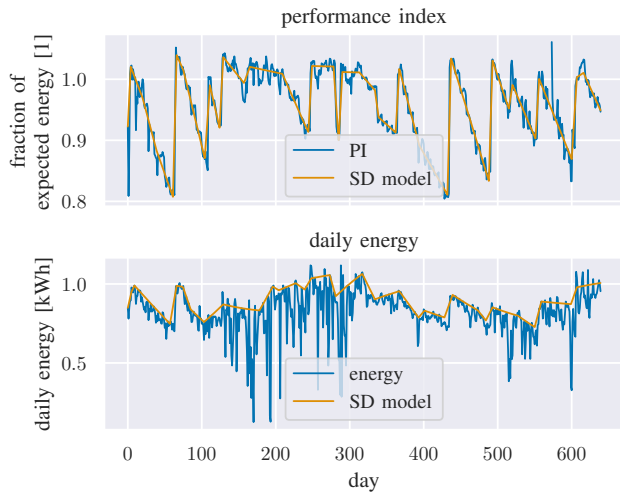}
}
\caption{PI and energy signals for the labeled soiling test system, with the 
denoised SD estimates overlaid in orange.}
\label{f-pv-data-after-SD}
\end{figure}

\begin{figure}
\centering
\resizebox{\columnwidth}{!}{
\import{figs/}{labeled-soiling-loss.pgf}
}
\caption{Comparison of the estimated soiling loss component from the PI signal 
and the unlabeled energy signal.}
\label{f-label-compare}
\end{figure}
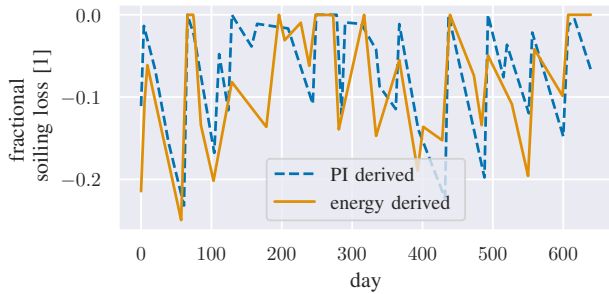

The labeled performance index and corresponding unlabeled energy signal was 
previously shown in figure~\ref{f-labeled-unlabeled}. In 
figure~\ref{f-pv-data-after-SD}, these signals are overlaid with the 
``denoised'' SD signal estimate, \ie, the sum of the estimated components 
excluding the first residual term. Note how the SD model for the PI assumes 
symmetric residuals ($\tau=0.5$) while the model for the energy signal assumes 
highly asymmetric residuals ($\tau=0.85$).
Figure~\ref{f-label-compare} shows the comparison between the soiling loss 
components estimated from the PI and from the unlabeled energy signal. Taking 
the loss component derived from the PI signal as groundtruth, we then calculate 
the three error metrics for the energy-derived soiling estimate. The loss MAE 
is 0.042558; the rate MAE is 0.004734, and the filtered rate MAE is 0.001756.

We find that the analysis of the unlabeled energy produces an estimate of 
soiling losses that agrees well with the PI-derived estimate. We observe that 
the unlabled analysis does worst around days 125--200, which was particularly 
rainy and cloudy. This lack of clear sky baseline during this period is seen to 
negatively impact the soiling estimate. However, with additional years of data, 
this may be improved due to the seasonal structure in the second component.


\subsection{Unlabeled data}

In~\cite{pvsc-shade}, we present a preliminary shade loss analysis of an 
unlabeled rooftop PV data set. As described in that manuscript, the shade 
analysis depends on first estimating and correcting the soiling losses in the 
signal. A view of the soiling signal decomposition for one system in this data 
set is shown in 
figure~\ref{f-soil-decomp}, and the soiling component in isolation is shown in 
figure~\ref{f-soil-isolate}.

\begin{figure}
\centering
\import{figs/}{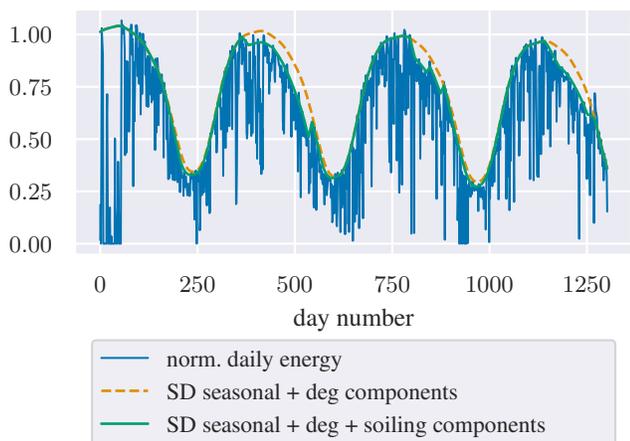}
\caption{An illustration of the soiling decomposition results for the unlabeled 
rooftop PV data discussed in~\cite{pvsc-shade}. The soiling trend is estimated 
to be the difference between the orange and green trends.}
\label{f-soil-decomp}
\end{figure}
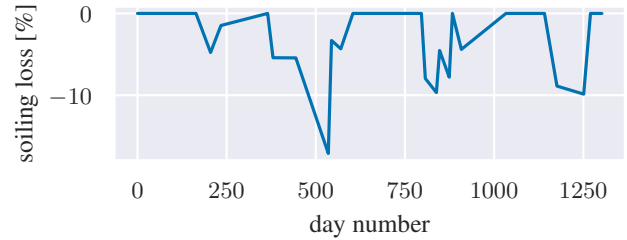
\begin{figure}
\centering
\import{figs/}{soil-isolate.pgf}
\caption{The isolated soiling trend for the unlabeled data analysis.}
\label{f-soil-isolate}
\end{figure}

\section{Conclusions}

We present a methodology, based on the signal decomposition (SD) framework, for 
estimating soiling losses PV system production data, \ie, time series 
measurements of generated real power or energy, typically over multiple years. 
Unique to this work is the capability to estimate soiling losses in raw, 
unlabeled power/energy data, rather than requiring a performance index. By 
utilizing the extensibility of the SD, we are able to design a signal 
decomposition model that is bespoke to the problem of estimating soiling 
losses. This results in a robust model for the soiling loss component, $x^4$, 
as well as adjustable residual component, $x^1$, that is able to model both 
unlabeled ($\tau=0.85$) and labeled ($\tau=0.5$) data. The ability to analyze 
unlabled PV data potentially unlocks huge potential in the form of fleet-scale 
datasets of hetergenous, distributed PV systems, which typically have 
internet-connected power electronics which generate time series of real power, 
but lack correlated meteorological measurements and possibly accurate system 
models. A software implementation is available in the Solar Data Tools 
package~\cite{Meyers2020b,solar-data-tools} and a demonstration notebook of the 
code usage is available online~\cite{soiling-notebook}.

\section*{Acknowledgments}
The author would like to thank 
Stephen Boyd, Justin Luke, Elsa Kam-Lum, Mayank Malik, and the entire GISMo 
Team at SLAC National 
Accelerator Laboratory for their input and feedback on this work. I also 
recognize the Seaborn plotting package for Python, which 
made the figures possible~\cite{Waskom2021}.

\small
\bibliographystyle{IEEEtran}
\bibliography{soiling2022}

\end{document}

%% file: defs.tex
\newcommand{\reals}{{\mbox{\bf R}}}

\newcommand{\eg}{{\it e.g.}}
\newcommand{\ie}{{\it i.e.}}

\newcommand{\BEAS}{\begin{eqnarray*}}
\newcommand{\EEAS}{\end{eqnarray*}}
\newcommand{\BEA}{\begin{eqnarray}}
\newcommand{\EEA}{\end{eqnarray}}
\newcommand{\BEQ}{\begin{equation}}
\newcommand{\EEQ}{\end{equation}}
\newcommand{\BIT}{\begin{itemize}}
\newcommand{\EIT}{\end{itemize}}

\newcounter{algorithmctr}[section]
\renewcommand{\thealgorithmctr}{\thesection.\arabic{algorithmctr}}
   {\refstepcounter{algorithmctr}\begin{list}{}{%
       \setlength{\rightmargin}{0\linewidth}%
       \setlength{\leftmargin}{.05\linewidth}}%
       \rmfamily\small
       \item[]{\setlength{\parskip}{0ex}\hrulefill\par%
        \nopagebreak{\bfseries\textsf{Algorithm \thealgorithmctr~}}}}%
   {{\setlength{\parskip}{-1ex}\nopagebreak\par\hrulefill} \end{list}}

%% file: figs/labeled-soiling-loss.pgf
\begingroup%
\makeatletter%
\begin{pgfpicture}%
\pgfpathrectangle{\pgfpointorigin}{\pgfqpoint{5.000000in}{2.500000in}}%
\pgfusepath{use as bounding box, clip}%
\begin{pgfscope}%
\pgfsetbuttcap%
\pgfsetmiterjoin%
\pgfsetlinewidth{0.000000pt}%
\definecolor{currentstroke}{rgb}{1.000000,1.000000,1.000000}%
\pgfsetstrokecolor{currentstroke}%
\pgfsetstrokeopacity{0.000000}%
\pgfsetdash{}{0pt}%
\pgfpathmoveto{\pgfqpoint{0.000000in}{0.000000in}}%
\pgfpathlineto{\pgfqpoint{5.000000in}{0.000000in}}%
\pgfpathlineto{\pgfqpoint{5.000000in}{2.500000in}}%
\pgfpathlineto{\pgfqpoint{0.000000in}{2.500000in}}%
\pgfpathlineto{\pgfqpoint{0.000000in}{0.000000in}}%
\pgfpathclose%
\pgfusepath{}%
\end{pgfscope}%
\begin{pgfscope}%
\pgfsetbuttcap%
\pgfsetmiterjoin%
\definecolor{currentfill}{rgb}{0.917647,0.917647,0.949020}%
\pgfsetfillcolor{currentfill}%
\pgfsetlinewidth{0.000000pt}%
\definecolor{currentstroke}{rgb}{0.000000,0.000000,0.000000}%
\pgfsetstrokecolor{currentstroke}%
\pgfsetstrokeopacity{0.000000}%
\pgfsetdash{}{0pt}%
\pgfpathmoveto{\pgfqpoint{1.065972in}{0.620833in}}%
\pgfpathlineto{\pgfqpoint{4.850000in}{0.620833in}}%
\pgfpathlineto{\pgfqpoint{4.850000in}{2.350000in}}%
\pgfpathlineto{\pgfqpoint{1.065972in}{2.350000in}}%
\pgfpathlineto{\pgfqpoint{1.065972in}{0.620833in}}%
\pgfpathclose%
\pgfusepath{fill}%
\end{pgfscope}%
\begin{pgfscope}%
\pgfpathrectangle{\pgfqpoint{1.065972in}{0.620833in}}{\pgfqpoint{3.784028in}{1.729167in}}%
\pgfusepath{clip}%
\pgfsetroundcap%
\pgfsetroundjoin%
\pgfsetlinewidth{1.003750pt}%
\definecolor{currentstroke}{rgb}{1.000000,1.000000,1.000000}%
\pgfsetstrokecolor{currentstroke}%
\pgfsetdash{}{0pt}%
\pgfpathmoveto{\pgfqpoint{1.237973in}{0.620833in}}%
\pgfpathlineto{\pgfqpoint{1.237973in}{2.350000in}}%
\pgfusepath{stroke}%
\end{pgfscope}%
\begin{pgfscope}%
\definecolor{textcolor}{rgb}{0.150000,0.150000,0.150000}%
\pgfsetstrokecolor{textcolor}%
\pgfsetfillcolor{textcolor}%
\pgftext[x=1.237973in,y=0.488888in,,top]{\color{textcolor}\rmfamily\fontsize{11.000000}{13.200000}\selectfont \(\displaystyle {0}\)}%
\end{pgfscope}%
\begin{pgfscope}%
\pgfpathrectangle{\pgfqpoint{1.065972in}{0.620833in}}{\pgfqpoint{3.784028in}{1.729167in}}%
\pgfusepath{clip}%
\pgfsetroundcap%
\pgfsetroundjoin%
\pgfsetlinewidth{1.003750pt}%
\definecolor{currentstroke}{rgb}{1.000000,1.000000,1.000000}%
\pgfsetstrokecolor{currentstroke}%
\pgfsetdash{}{0pt}%
\pgfpathmoveto{\pgfqpoint{1.776319in}{0.620833in}}%
\pgfpathlineto{\pgfqpoint{1.776319in}{2.350000in}}%
\pgfusepath{stroke}%
\end{pgfscope}%
\begin{pgfscope}%
\definecolor{textcolor}{rgb}{0.150000,0.150000,0.150000}%
\pgfsetstrokecolor{textcolor}%
\pgfsetfillcolor{textcolor}%
\pgftext[x=1.776319in,y=0.488888in,,top]{\color{textcolor}\rmfamily\fontsize{11.000000}{13.200000}\selectfont \(\displaystyle {100}\)}%
\end{pgfscope}%
\begin{pgfscope}%
\pgfpathrectangle{\pgfqpoint{1.065972in}{0.620833in}}{\pgfqpoint{3.784028in}{1.729167in}}%
\pgfusepath{clip}%
\pgfsetroundcap%
\pgfsetroundjoin%
\pgfsetlinewidth{1.003750pt}%
\definecolor{currentstroke}{rgb}{1.000000,1.000000,1.000000}%
\pgfsetstrokecolor{currentstroke}%
\pgfsetdash{}{0pt}%
\pgfpathmoveto{\pgfqpoint{2.314664in}{0.620833in}}%
\pgfpathlineto{\pgfqpoint{2.314664in}{2.350000in}}%
\pgfusepath{stroke}%
\end{pgfscope}%
\begin{pgfscope}%
\definecolor{textcolor}{rgb}{0.150000,0.150000,0.150000}%
\pgfsetstrokecolor{textcolor}%
\pgfsetfillcolor{textcolor}%
\pgftext[x=2.314664in,y=0.488888in,,top]{\color{textcolor}\rmfamily\fontsize{11.000000}{13.200000}\selectfont \(\displaystyle {200}\)}%
\end{pgfscope}%
\begin{pgfscope}%
\pgfpathrectangle{\pgfqpoint{1.065972in}{0.620833in}}{\pgfqpoint{3.784028in}{1.729167in}}%
\pgfusepath{clip}%
\pgfsetroundcap%
\pgfsetroundjoin%
\pgfsetlinewidth{1.003750pt}%
\definecolor{currentstroke}{rgb}{1.000000,1.000000,1.000000}%
\pgfsetstrokecolor{currentstroke}%
\pgfsetdash{}{0pt}%
\pgfpathmoveto{\pgfqpoint{2.853009in}{0.620833in}}%
\pgfpathlineto{\pgfqpoint{2.853009in}{2.350000in}}%
\pgfusepath{stroke}%
\end{pgfscope}%
\begin{pgfscope}%
\definecolor{textcolor}{rgb}{0.150000,0.150000,0.150000}%
\pgfsetstrokecolor{textcolor}%
\pgfsetfillcolor{textcolor}%
\pgftext[x=2.853009in,y=0.488888in,,top]{\color{textcolor}\rmfamily\fontsize{11.000000}{13.200000}\selectfont \(\displaystyle {300}\)}%
\end{pgfscope}%
\begin{pgfscope}%
\pgfpathrectangle{\pgfqpoint{1.065972in}{0.620833in}}{\pgfqpoint{3.784028in}{1.729167in}}%
\pgfusepath{clip}%
\pgfsetroundcap%
\pgfsetroundjoin%
\pgfsetlinewidth{1.003750pt}%
\definecolor{currentstroke}{rgb}{1.000000,1.000000,1.000000}%
\pgfsetstrokecolor{currentstroke}%
\pgfsetdash{}{0pt}%
\pgfpathmoveto{\pgfqpoint{3.391354in}{0.620833in}}%
\pgfpathlineto{\pgfqpoint{3.391354in}{2.350000in}}%
\pgfusepath{stroke}%
\end{pgfscope}%
\begin{pgfscope}%
\definecolor{textcolor}{rgb}{0.150000,0.150000,0.150000}%
\pgfsetstrokecolor{textcolor}%
\pgfsetfillcolor{textcolor}%
\pgftext[x=3.391354in,y=0.488888in,,top]{\color{textcolor}\rmfamily\fontsize{11.000000}{13.200000}\selectfont \(\displaystyle {400}\)}%
\end{pgfscope}%
\begin{pgfscope}%
\pgfpathrectangle{\pgfqpoint{1.065972in}{0.620833in}}{\pgfqpoint{3.784028in}{1.729167in}}%
\pgfusepath{clip}%
\pgfsetroundcap%
\pgfsetroundjoin%
\pgfsetlinewidth{1.003750pt}%
\definecolor{currentstroke}{rgb}{1.000000,1.000000,1.000000}%
\pgfsetstrokecolor{currentstroke}%
\pgfsetdash{}{0pt}%
\pgfpathmoveto{\pgfqpoint{3.929699in}{0.620833in}}%
\pgfpathlineto{\pgfqpoint{3.929699in}{2.350000in}}%
\pgfusepath{stroke}%
\end{pgfscope}%
\begin{pgfscope}%
\definecolor{textcolor}{rgb}{0.150000,0.150000,0.150000}%
\pgfsetstrokecolor{textcolor}%
\pgfsetfillcolor{textcolor}%
\pgftext[x=3.929699in,y=0.488888in,,top]{\color{textcolor}\rmfamily\fontsize{11.000000}{13.200000}\selectfont \(\displaystyle {500}\)}%
\end{pgfscope}%
\begin{pgfscope}%
\pgfpathrectangle{\pgfqpoint{1.065972in}{0.620833in}}{\pgfqpoint{3.784028in}{1.729167in}}%
\pgfusepath{clip}%
\pgfsetroundcap%
\pgfsetroundjoin%
\pgfsetlinewidth{1.003750pt}%
\definecolor{currentstroke}{rgb}{1.000000,1.000000,1.000000}%
\pgfsetstrokecolor{currentstroke}%
\pgfsetdash{}{0pt}%
\pgfpathmoveto{\pgfqpoint{4.468044in}{0.620833in}}%
\pgfpathlineto{\pgfqpoint{4.468044in}{2.350000in}}%
\pgfusepath{stroke}%
\end{pgfscope}%
\begin{pgfscope}%
\definecolor{textcolor}{rgb}{0.150000,0.150000,0.150000}%
\pgfsetstrokecolor{textcolor}%
\pgfsetfillcolor{textcolor}%
\pgftext[x=4.468044in,y=0.488888in,,top]{\color{textcolor}\rmfamily\fontsize{11.000000}{13.200000}\selectfont \(\displaystyle {600}\)}%
\end{pgfscope}%
\begin{pgfscope}%
\definecolor{textcolor}{rgb}{0.150000,0.150000,0.150000}%
\pgfsetstrokecolor{textcolor}%
\pgfsetfillcolor{textcolor}%
\pgftext[x=2.957986in,y=0.298147in,,top]{\color{textcolor}\rmfamily\fontsize{12.000000}{14.400000}\selectfont day}%
\end{pgfscope}%
\begin{pgfscope}%
\pgfpathrectangle{\pgfqpoint{1.065972in}{0.620833in}}{\pgfqpoint{3.784028in}{1.729167in}}%
\pgfusepath{clip}%
\pgfsetroundcap%
\pgfsetroundjoin%
\pgfsetlinewidth{1.003750pt}%
\definecolor{currentstroke}{rgb}{1.000000,1.000000,1.000000}%
\pgfsetstrokecolor{currentstroke}%
\pgfsetdash{}{0pt}%
\pgfpathmoveto{\pgfqpoint{1.065972in}{1.011225in}}%
\pgfpathlineto{\pgfqpoint{4.850000in}{1.011225in}}%
\pgfusepath{stroke}%
\end{pgfscope}%
\begin{pgfscope}%
\definecolor{textcolor}{rgb}{0.150000,0.150000,0.150000}%
\pgfsetstrokecolor{textcolor}%
\pgfsetfillcolor{textcolor}%
\pgftext[x=0.621411in, y=0.958418in, left, base]{\color{textcolor}\rmfamily\fontsize{11.000000}{13.200000}\selectfont \(\displaystyle {-0.2}\)}%
\end{pgfscope}%
\begin{pgfscope}%
\pgfpathrectangle{\pgfqpoint{1.065972in}{0.620833in}}{\pgfqpoint{3.784028in}{1.729167in}}%
\pgfusepath{clip}%
\pgfsetroundcap%
\pgfsetroundjoin%
\pgfsetlinewidth{1.003750pt}%
\definecolor{currentstroke}{rgb}{1.000000,1.000000,1.000000}%
\pgfsetstrokecolor{currentstroke}%
\pgfsetdash{}{0pt}%
\pgfpathmoveto{\pgfqpoint{1.065972in}{1.641313in}}%
\pgfpathlineto{\pgfqpoint{4.850000in}{1.641313in}}%
\pgfusepath{stroke}%
\end{pgfscope}%
\begin{pgfscope}%
\definecolor{textcolor}{rgb}{0.150000,0.150000,0.150000}%
\pgfsetstrokecolor{textcolor}%
\pgfsetfillcolor{textcolor}%
\pgftext[x=0.621411in, y=1.588506in, left, base]{\color{textcolor}\rmfamily\fontsize{11.000000}{13.200000}\selectfont \(\displaystyle {-0.1}\)}%
\end{pgfscope}%
\begin{pgfscope}%
\pgfpathrectangle{\pgfqpoint{1.065972in}{0.620833in}}{\pgfqpoint{3.784028in}{1.729167in}}%
\pgfusepath{clip}%
\pgfsetroundcap%
\pgfsetroundjoin%
\pgfsetlinewidth{1.003750pt}%
\definecolor{currentstroke}{rgb}{1.000000,1.000000,1.000000}%
\pgfsetstrokecolor{currentstroke}%
\pgfsetdash{}{0pt}%
\pgfpathmoveto{\pgfqpoint{1.065972in}{2.271401in}}%
\pgfpathlineto{\pgfqpoint{4.850000in}{2.271401in}}%
\pgfusepath{stroke}%
\end{pgfscope}%
\begin{pgfscope}%
\definecolor{textcolor}{rgb}{0.150000,0.150000,0.150000}%
\pgfsetstrokecolor{textcolor}%
\pgfsetfillcolor{textcolor}%
\pgftext[x=0.739699in, y=2.218595in, left, base]{\color{textcolor}\rmfamily\fontsize{11.000000}{13.200000}\selectfont \(\displaystyle {0.0}\)}%
\end{pgfscope}%
\begin{pgfscope}%
\definecolor{textcolor}{rgb}{0.150000,0.150000,0.150000}%
\pgfsetstrokecolor{textcolor}%
\pgfsetfillcolor{textcolor}%
\pgftext[x=0.341782in, y=1.147698in, left, base,rotate=90.000000]{\color{textcolor}\rmfamily\fontsize{12.000000}{14.400000}\selectfont fractional}%
\end{pgfscope}%
\begin{pgfscope}%
\definecolor{textcolor}{rgb}{0.150000,0.150000,0.150000}%
\pgfsetstrokecolor{textcolor}%
\pgfsetfillcolor{textcolor}%
\pgftext[x=0.524189in, y=0.989945in, left, base,rotate=90.000000]{\color{textcolor}\rmfamily\fontsize{12.000000}{14.400000}\selectfont soiling loss [1]}%
\end{pgfscope}%
\begin{pgfscope}%
\pgfpathrectangle{\pgfqpoint{1.065972in}{0.620833in}}{\pgfqpoint{3.784028in}{1.729167in}}%
\pgfusepath{clip}%
\pgfsetbuttcap%
\pgfsetroundjoin%
\pgfsetlinewidth{1.505625pt}%
\definecolor{currentstroke}{rgb}{0.003922,0.450980,0.698039}%
\pgfsetstrokecolor{currentstroke}%
\pgfsetdash{{5.550000pt}{2.400000pt}}{0.000000pt}%
\pgfpathmoveto{\pgfqpoint{1.237973in}{1.572125in}}%
\pgfpathlineto{\pgfqpoint{1.259507in}{2.186429in}}%
\pgfpathlineto{\pgfqpoint{1.345643in}{1.852410in}}%
\pgfpathlineto{\pgfqpoint{1.447928in}{1.295377in}}%
\pgfpathlineto{\pgfqpoint{1.566364in}{0.811041in}}%
\pgfpathlineto{\pgfqpoint{1.593281in}{2.271401in}}%
\pgfpathlineto{\pgfqpoint{1.641732in}{2.116489in}}%
\pgfpathlineto{\pgfqpoint{1.797852in}{1.214209in}}%
\pgfpathlineto{\pgfqpoint{1.835537in}{1.970339in}}%
\pgfpathlineto{\pgfqpoint{1.905521in}{1.540200in}}%
\pgfpathlineto{\pgfqpoint{1.932439in}{2.271401in}}%
\pgfpathlineto{\pgfqpoint{2.083175in}{2.028584in}}%
\pgfpathlineto{\pgfqpoint{2.126243in}{2.202063in}}%
\pgfpathlineto{\pgfqpoint{2.363115in}{2.166847in}}%
\pgfpathlineto{\pgfqpoint{2.551536in}{1.588595in}}%
\pgfpathlineto{\pgfqpoint{2.578453in}{2.271400in}}%
\pgfpathlineto{\pgfqpoint{2.734573in}{2.271401in}}%
\pgfpathlineto{\pgfqpoint{2.772257in}{1.510244in}}%
\pgfpathlineto{\pgfqpoint{2.799174in}{2.213193in}}%
\pgfpathlineto{\pgfqpoint{2.922994in}{2.201215in}}%
\pgfpathlineto{\pgfqpoint{3.036046in}{2.005121in}}%
\pgfpathlineto{\pgfqpoint{3.062963in}{1.723467in}}%
\pgfpathlineto{\pgfqpoint{3.186783in}{1.549062in}}%
\pgfpathlineto{\pgfqpoint{3.213700in}{2.200755in}}%
\pgfpathlineto{\pgfqpoint{3.359053in}{1.392508in}}%
\pgfpathlineto{\pgfqpoint{3.563624in}{0.865969in}}%
\pgfpathlineto{\pgfqpoint{3.590542in}{2.271401in}}%
\pgfpathlineto{\pgfqpoint{3.865098in}{1.026190in}}%
\pgfpathlineto{\pgfqpoint{3.892015in}{2.271401in}}%
\pgfpathlineto{\pgfqpoint{4.010451in}{1.793455in}}%
\pgfpathlineto{\pgfqpoint{4.037368in}{2.042239in}}%
\pgfpathlineto{\pgfqpoint{4.204255in}{1.517520in}}%
\pgfpathlineto{\pgfqpoint{4.231172in}{2.135499in}}%
\pgfpathlineto{\pgfqpoint{4.468044in}{1.340371in}}%
\pgfpathlineto{\pgfqpoint{4.505728in}{2.187771in}}%
\pgfpathlineto{\pgfqpoint{4.554179in}{2.237206in}}%
\pgfpathlineto{\pgfqpoint{4.677999in}{1.853392in}}%
\pgfpathlineto{\pgfqpoint{4.677999in}{1.853392in}}%
\pgfusepath{stroke}%
\end{pgfscope}%
\begin{pgfscope}%
\pgfpathrectangle{\pgfqpoint{1.065972in}{0.620833in}}{\pgfqpoint{3.784028in}{1.729167in}}%
\pgfusepath{clip}%
\pgfsetroundcap%
\pgfsetroundjoin%
\pgfsetlinewidth{1.505625pt}%
\definecolor{currentstroke}{rgb}{0.870588,0.560784,0.019608}%
\pgfsetstrokecolor{currentstroke}%
\pgfsetdash{}{0pt}%
\pgfpathmoveto{\pgfqpoint{1.237973in}{0.920504in}}%
\pgfpathlineto{\pgfqpoint{1.264891in}{1.604027in}}%
\pgfpathlineto{\pgfqpoint{1.286425in}{1.886192in}}%
\pgfpathlineto{\pgfqpoint{1.544830in}{0.699431in}}%
\pgfpathlineto{\pgfqpoint{1.593281in}{2.271401in}}%
\pgfpathlineto{\pgfqpoint{1.636349in}{2.271401in}}%
\pgfpathlineto{\pgfqpoint{1.641732in}{2.232599in}}%
\pgfpathlineto{\pgfqpoint{1.695567in}{1.427644in}}%
\pgfpathlineto{\pgfqpoint{1.792469in}{1.000295in}}%
\pgfpathlineto{\pgfqpoint{1.932439in}{1.756169in}}%
\pgfpathlineto{\pgfqpoint{2.196228in}{1.413001in}}%
\pgfpathlineto{\pgfqpoint{2.293130in}{2.271401in}}%
\pgfpathlineto{\pgfqpoint{2.336198in}{2.076646in}}%
\pgfpathlineto{\pgfqpoint{2.460017in}{2.210815in}}%
\pgfpathlineto{\pgfqpoint{2.524618in}{1.880724in}}%
\pgfpathlineto{\pgfqpoint{2.573069in}{2.271401in}}%
\pgfpathlineto{\pgfqpoint{2.707656in}{2.269770in}}%
\pgfpathlineto{\pgfqpoint{2.750723in}{1.393752in}}%
\pgfpathlineto{\pgfqpoint{2.944527in}{2.271401in}}%
\pgfpathlineto{\pgfqpoint{3.036046in}{1.343739in}}%
\pgfpathlineto{\pgfqpoint{3.213700in}{1.924005in}}%
\pgfpathlineto{\pgfqpoint{3.353670in}{1.078927in}}%
\pgfpathlineto{\pgfqpoint{3.396737in}{1.414097in}}%
\pgfpathlineto{\pgfqpoint{3.542091in}{1.310409in}}%
\pgfpathlineto{\pgfqpoint{3.601309in}{2.271401in}}%
\pgfpathlineto{\pgfqpoint{3.784346in}{1.806469in}}%
\pgfpathlineto{\pgfqpoint{3.843564in}{1.426587in}}%
\pgfpathlineto{\pgfqpoint{3.892015in}{1.960076in}}%
\pgfpathlineto{\pgfqpoint{4.075052in}{1.586978in}}%
\pgfpathlineto{\pgfqpoint{4.198872in}{1.037270in}}%
\pgfpathlineto{\pgfqpoint{4.247323in}{2.007928in}}%
\pgfpathlineto{\pgfqpoint{4.462661in}{1.650442in}}%
\pgfpathlineto{\pgfqpoint{4.505728in}{2.271401in}}%
\pgfpathlineto{\pgfqpoint{4.677999in}{2.271401in}}%
\pgfpathlineto{\pgfqpoint{4.677999in}{2.271401in}}%
\pgfusepath{stroke}%
\end{pgfscope}%
\begin{pgfscope}%
\pgfsetrectcap%
\pgfsetmiterjoin%
\pgfsetlinewidth{1.254687pt}%
\definecolor{currentstroke}{rgb}{1.000000,1.000000,1.000000}%
\pgfsetstrokecolor{currentstroke}%
\pgfsetdash{}{0pt}%
\pgfpathmoveto{\pgfqpoint{1.065972in}{0.620833in}}%
\pgfpathlineto{\pgfqpoint{1.065972in}{2.350000in}}%
\pgfusepath{stroke}%
\end{pgfscope}%
\begin{pgfscope}%
\pgfsetrectcap%
\pgfsetmiterjoin%
\pgfsetlinewidth{1.254687pt}%
\definecolor{currentstroke}{rgb}{1.000000,1.000000,1.000000}%
\pgfsetstrokecolor{currentstroke}%
\pgfsetdash{}{0pt}%
\pgfpathmoveto{\pgfqpoint{4.850000in}{0.620833in}}%
\pgfpathlineto{\pgfqpoint{4.850000in}{2.350000in}}%
\pgfusepath{stroke}%
\end{pgfscope}%
\begin{pgfscope}%
\pgfsetrectcap%
\pgfsetmiterjoin%
\pgfsetlinewidth{1.254687pt}%
\definecolor{currentstroke}{rgb}{1.000000,1.000000,1.000000}%
\pgfsetstrokecolor{currentstroke}%
\pgfsetdash{}{0pt}%
\pgfpathmoveto{\pgfqpoint{1.065972in}{0.620833in}}%
\pgfpathlineto{\pgfqpoint{4.850000in}{0.620833in}}%
\pgfusepath{stroke}%
\end{pgfscope}%
\begin{pgfscope}%
\pgfsetrectcap%
\pgfsetmiterjoin%
\pgfsetlinewidth{1.254687pt}%
\definecolor{currentstroke}{rgb}{1.000000,1.000000,1.000000}%
\pgfsetstrokecolor{currentstroke}%
\pgfsetdash{}{0pt}%
\pgfpathmoveto{\pgfqpoint{1.065972in}{2.350000in}}%
\pgfpathlineto{\pgfqpoint{4.850000in}{2.350000in}}%
\pgfusepath{stroke}%
\end{pgfscope}%
\begin{pgfscope}%
\pgfsetbuttcap%
\pgfsetmiterjoin%
\definecolor{currentfill}{rgb}{0.917647,0.917647,0.949020}%
\pgfsetfillcolor{currentfill}%
\pgfsetfillopacity{0.800000}%
\pgfsetlinewidth{1.003750pt}%
\definecolor{currentstroke}{rgb}{0.800000,0.800000,0.800000}%
\pgfsetstrokecolor{currentstroke}%
\pgfsetstrokeopacity{0.800000}%
\pgfsetdash{}{0pt}%
\pgfpathmoveto{\pgfqpoint{2.229409in}{0.697222in}}%
\pgfpathlineto{\pgfqpoint{3.686563in}{0.697222in}}%
\pgfpathquadraticcurveto{\pgfqpoint{3.717119in}{0.697222in}}{\pgfqpoint{3.717119in}{0.727777in}}%
\pgfpathlineto{\pgfqpoint{3.717119in}{1.138310in}}%
\pgfpathquadraticcurveto{\pgfqpoint{3.717119in}{1.168865in}}{\pgfqpoint{3.686563in}{1.168865in}}%
\pgfpathlineto{\pgfqpoint{2.229409in}{1.168865in}}%
\pgfpathquadraticcurveto{\pgfqpoint{2.198854in}{1.168865in}}{\pgfqpoint{2.198854in}{1.138310in}}%
\pgfpathlineto{\pgfqpoint{2.198854in}{0.727777in}}%
\pgfpathquadraticcurveto{\pgfqpoint{2.198854in}{0.697222in}}{\pgfqpoint{2.229409in}{0.697222in}}%
\pgfpathlineto{\pgfqpoint{2.229409in}{0.697222in}}%
\pgfpathclose%
\pgfusepath{stroke,fill}%
\end{pgfscope}%
\begin{pgfscope}%
\pgfsetbuttcap%
\pgfsetroundjoin%
\pgfsetlinewidth{1.505625pt}%
\definecolor{currentstroke}{rgb}{0.003922,0.450980,0.698039}%
\pgfsetstrokecolor{currentstroke}%
\pgfsetdash{{5.550000pt}{2.400000pt}}{0.000000pt}%
\pgfpathmoveto{\pgfqpoint{2.259965in}{1.054282in}}%
\pgfpathlineto{\pgfqpoint{2.412742in}{1.054282in}}%
\pgfpathlineto{\pgfqpoint{2.565520in}{1.054282in}}%
\pgfusepath{stroke}%
\end{pgfscope}%
\begin{pgfscope}%
\definecolor{textcolor}{rgb}{0.150000,0.150000,0.150000}%
\pgfsetstrokecolor{textcolor}%
\pgfsetfillcolor{textcolor}%
\pgftext[x=2.687742in,y=1.000810in,left,base]{\color{textcolor}\rmfamily\fontsize{11.000000}{13.200000}\selectfont PI derived}%
\end{pgfscope}%
\begin{pgfscope}%
\pgfsetroundcap%
\pgfsetroundjoin%
\pgfsetlinewidth{1.505625pt}%
\definecolor{currentstroke}{rgb}{0.870588,0.560784,0.019608}%
\pgfsetstrokecolor{currentstroke}%
\pgfsetdash{}{0pt}%
\pgfpathmoveto{\pgfqpoint{2.259965in}{0.841377in}}%
\pgfpathlineto{\pgfqpoint{2.412742in}{0.841377in}}%
\pgfpathlineto{\pgfqpoint{2.565520in}{0.841377in}}%
\pgfusepath{stroke}%
\end{pgfscope}%
\begin{pgfscope}%
\definecolor{textcolor}{rgb}{0.150000,0.150000,0.150000}%
\pgfsetstrokecolor{textcolor}%
\pgfsetfillcolor{textcolor}%
\pgftext[x=2.687742in,y=0.787904in,left,base]{\color{textcolor}\rmfamily\fontsize{11.000000}{13.200000}\selectfont energy derived}%
\end{pgfscope}%
\end{pgfpicture}%
\makeatother%
\endgroup%

%% file: figs/soil-isolate.pgf
\begingroup%
\makeatletter%
\begin{pgfpicture}%
\pgfpathrectangle{\pgfpointorigin}{\pgfqpoint{3.500000in}{1.500000in}}%
\pgfusepath{use as bounding box, clip}%
\begin{pgfscope}%
\pgfsetbuttcap%
\pgfsetmiterjoin%
\definecolor{currentfill}{rgb}{1.000000,1.000000,1.000000}%
\pgfsetfillcolor{currentfill}%
\pgfsetlinewidth{0.000000pt}%
\definecolor{currentstroke}{rgb}{1.000000,1.000000,1.000000}%
\pgfsetstrokecolor{currentstroke}%
\pgfsetstrokeopacity{0.000000}%
\pgfsetdash{}{0pt}%
\pgfpathmoveto{\pgfqpoint{0.000000in}{0.000000in}}%
\pgfpathlineto{\pgfqpoint{3.500000in}{0.000000in}}%
\pgfpathlineto{\pgfqpoint{3.500000in}{1.500000in}}%
\pgfpathlineto{\pgfqpoint{0.000000in}{1.500000in}}%
\pgfpathlineto{\pgfqpoint{0.000000in}{0.000000in}}%
\pgfpathclose%
\pgfusepath{fill}%
\end{pgfscope}%
\begin{pgfscope}%
\pgfsetbuttcap%
\pgfsetmiterjoin%
\definecolor{currentfill}{rgb}{0.917647,0.917647,0.949020}%
\pgfsetfillcolor{currentfill}%
\pgfsetlinewidth{0.000000pt}%
\definecolor{currentstroke}{rgb}{0.000000,0.000000,0.000000}%
\pgfsetstrokecolor{currentstroke}%
\pgfsetstrokeopacity{0.000000}%
\pgfsetdash{}{0pt}%
\pgfpathmoveto{\pgfqpoint{0.682116in}{0.549401in}}%
\pgfpathlineto{\pgfqpoint{3.356000in}{0.549401in}}%
\pgfpathlineto{\pgfqpoint{3.356000in}{1.356000in}}%
\pgfpathlineto{\pgfqpoint{0.682116in}{1.356000in}}%
\pgfpathlineto{\pgfqpoint{0.682116in}{0.549401in}}%
\pgfpathclose%
\pgfusepath{fill}%
\end{pgfscope}%
\begin{pgfscope}%
\pgfpathrectangle{\pgfqpoint{0.682116in}{0.549401in}}{\pgfqpoint{2.673884in}{0.806599in}}%
\pgfusepath{clip}%
\pgfsetroundcap%
\pgfsetroundjoin%
\pgfsetlinewidth{0.803000pt}%
\definecolor{currentstroke}{rgb}{1.000000,1.000000,1.000000}%
\pgfsetstrokecolor{currentstroke}%
\pgfsetdash{}{0pt}%
\pgfpathmoveto{\pgfqpoint{0.803656in}{0.549401in}}%
\pgfpathlineto{\pgfqpoint{0.803656in}{1.356000in}}%
\pgfusepath{stroke}%
\end{pgfscope}%
\begin{pgfscope}%
\definecolor{textcolor}{rgb}{0.150000,0.150000,0.150000}%
\pgfsetstrokecolor{textcolor}%
\pgfsetfillcolor{textcolor}%
\pgftext[x=0.803656in,y=0.434123in,,top]{\color{textcolor}\rmfamily\fontsize{8.800000}{10.560000}\selectfont \(\displaystyle {0}\)}%
\end{pgfscope}%
\begin{pgfscope}%
\pgfpathrectangle{\pgfqpoint{0.682116in}{0.549401in}}{\pgfqpoint{2.673884in}{0.806599in}}%
\pgfusepath{clip}%
\pgfsetroundcap%
\pgfsetroundjoin%
\pgfsetlinewidth{0.803000pt}%
\definecolor{currentstroke}{rgb}{1.000000,1.000000,1.000000}%
\pgfsetstrokecolor{currentstroke}%
\pgfsetdash{}{0pt}%
\pgfpathmoveto{\pgfqpoint{1.270400in}{0.549401in}}%
\pgfpathlineto{\pgfqpoint{1.270400in}{1.356000in}}%
\pgfusepath{stroke}%
\end{pgfscope}%
\begin{pgfscope}%
\definecolor{textcolor}{rgb}{0.150000,0.150000,0.150000}%
\pgfsetstrokecolor{textcolor}%
\pgfsetfillcolor{textcolor}%
\pgftext[x=1.270400in,y=0.434123in,,top]{\color{textcolor}\rmfamily\fontsize{8.800000}{10.560000}\selectfont \(\displaystyle {250}\)}%
\end{pgfscope}%
\begin{pgfscope}%
\pgfpathrectangle{\pgfqpoint{0.682116in}{0.549401in}}{\pgfqpoint{2.673884in}{0.806599in}}%
\pgfusepath{clip}%
\pgfsetroundcap%
\pgfsetroundjoin%
\pgfsetlinewidth{0.803000pt}%
\definecolor{currentstroke}{rgb}{1.000000,1.000000,1.000000}%
\pgfsetstrokecolor{currentstroke}%
\pgfsetdash{}{0pt}%
\pgfpathmoveto{\pgfqpoint{1.737144in}{0.549401in}}%
\pgfpathlineto{\pgfqpoint{1.737144in}{1.356000in}}%
\pgfusepath{stroke}%
\end{pgfscope}%
\begin{pgfscope}%
\definecolor{textcolor}{rgb}{0.150000,0.150000,0.150000}%
\pgfsetstrokecolor{textcolor}%
\pgfsetfillcolor{textcolor}%
\pgftext[x=1.737144in,y=0.434123in,,top]{\color{textcolor}\rmfamily\fontsize{8.800000}{10.560000}\selectfont \(\displaystyle {500}\)}%
\end{pgfscope}%
\begin{pgfscope}%
\pgfpathrectangle{\pgfqpoint{0.682116in}{0.549401in}}{\pgfqpoint{2.673884in}{0.806599in}}%
\pgfusepath{clip}%
\pgfsetroundcap%
\pgfsetroundjoin%
\pgfsetlinewidth{0.803000pt}%
\definecolor{currentstroke}{rgb}{1.000000,1.000000,1.000000}%
\pgfsetstrokecolor{currentstroke}%
\pgfsetdash{}{0pt}%
\pgfpathmoveto{\pgfqpoint{2.203889in}{0.549401in}}%
\pgfpathlineto{\pgfqpoint{2.203889in}{1.356000in}}%
\pgfusepath{stroke}%
\end{pgfscope}%
\begin{pgfscope}%
\definecolor{textcolor}{rgb}{0.150000,0.150000,0.150000}%
\pgfsetstrokecolor{textcolor}%
\pgfsetfillcolor{textcolor}%
\pgftext[x=2.203889in,y=0.434123in,,top]{\color{textcolor}\rmfamily\fontsize{8.800000}{10.560000}\selectfont \(\displaystyle {750}\)}%
\end{pgfscope}%
\begin{pgfscope}%
\pgfpathrectangle{\pgfqpoint{0.682116in}{0.549401in}}{\pgfqpoint{2.673884in}{0.806599in}}%
\pgfusepath{clip}%
\pgfsetroundcap%
\pgfsetroundjoin%
\pgfsetlinewidth{0.803000pt}%
\definecolor{currentstroke}{rgb}{1.000000,1.000000,1.000000}%
\pgfsetstrokecolor{currentstroke}%
\pgfsetdash{}{0pt}%
\pgfpathmoveto{\pgfqpoint{2.670633in}{0.549401in}}%
\pgfpathlineto{\pgfqpoint{2.670633in}{1.356000in}}%
\pgfusepath{stroke}%
\end{pgfscope}%
\begin{pgfscope}%
\definecolor{textcolor}{rgb}{0.150000,0.150000,0.150000}%
\pgfsetstrokecolor{textcolor}%
\pgfsetfillcolor{textcolor}%
\pgftext[x=2.670633in,y=0.434123in,,top]{\color{textcolor}\rmfamily\fontsize{8.800000}{10.560000}\selectfont \(\displaystyle {1000}\)}%
\end{pgfscope}%
\begin{pgfscope}%
\pgfpathrectangle{\pgfqpoint{0.682116in}{0.549401in}}{\pgfqpoint{2.673884in}{0.806599in}}%
\pgfusepath{clip}%
\pgfsetroundcap%
\pgfsetroundjoin%
\pgfsetlinewidth{0.803000pt}%
\definecolor{currentstroke}{rgb}{1.000000,1.000000,1.000000}%
\pgfsetstrokecolor{currentstroke}%
\pgfsetdash{}{0pt}%
\pgfpathmoveto{\pgfqpoint{3.137377in}{0.549401in}}%
\pgfpathlineto{\pgfqpoint{3.137377in}{1.356000in}}%
\pgfusepath{stroke}%
\end{pgfscope}%
\begin{pgfscope}%
\definecolor{textcolor}{rgb}{0.150000,0.150000,0.150000}%
\pgfsetstrokecolor{textcolor}%
\pgfsetfillcolor{textcolor}%
\pgftext[x=3.137377in,y=0.434123in,,top]{\color{textcolor}\rmfamily\fontsize{8.800000}{10.560000}\selectfont \(\displaystyle {1250}\)}%
\end{pgfscope}%
\begin{pgfscope}%
\definecolor{textcolor}{rgb}{0.150000,0.150000,0.150000}%
\pgfsetstrokecolor{textcolor}%
\pgfsetfillcolor{textcolor}%
\pgftext[x=2.019058in,y=0.267457in,,top]{\color{textcolor}\rmfamily\fontsize{9.600000}{11.520000}\selectfont day number}%
\end{pgfscope}%
\begin{pgfscope}%
\pgfpathrectangle{\pgfqpoint{0.682116in}{0.549401in}}{\pgfqpoint{2.673884in}{0.806599in}}%
\pgfusepath{clip}%
\pgfsetroundcap%
\pgfsetroundjoin%
\pgfsetlinewidth{0.803000pt}%
\definecolor{currentstroke}{rgb}{1.000000,1.000000,1.000000}%
\pgfsetstrokecolor{currentstroke}%
\pgfsetdash{}{0pt}%
\pgfpathmoveto{\pgfqpoint{0.682116in}{0.890834in}}%
\pgfpathlineto{\pgfqpoint{3.356000in}{0.890834in}}%
\pgfusepath{stroke}%
\end{pgfscope}%
\begin{pgfscope}%
\definecolor{textcolor}{rgb}{0.150000,0.150000,0.150000}%
\pgfsetstrokecolor{textcolor}%
\pgfsetfillcolor{textcolor}%
\pgftext[x=0.338444in, y=0.847432in, left, base]{\color{textcolor}\rmfamily\fontsize{8.800000}{10.560000}\selectfont \(\displaystyle {-10}\)}%
\end{pgfscope}%
\begin{pgfscope}%
\pgfpathrectangle{\pgfqpoint{0.682116in}{0.549401in}}{\pgfqpoint{2.673884in}{0.806599in}}%
\pgfusepath{clip}%
\pgfsetroundcap%
\pgfsetroundjoin%
\pgfsetlinewidth{0.803000pt}%
\definecolor{currentstroke}{rgb}{1.000000,1.000000,1.000000}%
\pgfsetstrokecolor{currentstroke}%
\pgfsetdash{}{0pt}%
\pgfpathmoveto{\pgfqpoint{0.682116in}{1.319336in}}%
\pgfpathlineto{\pgfqpoint{3.356000in}{1.319336in}}%
\pgfusepath{stroke}%
\end{pgfscope}%
\begin{pgfscope}%
\definecolor{textcolor}{rgb}{0.150000,0.150000,0.150000}%
\pgfsetstrokecolor{textcolor}%
\pgfsetfillcolor{textcolor}%
\pgftext[x=0.502602in, y=1.275934in, left, base]{\color{textcolor}\rmfamily\fontsize{8.800000}{10.560000}\selectfont \(\displaystyle {0}\)}%
\end{pgfscope}%
\begin{pgfscope}%
\definecolor{textcolor}{rgb}{0.150000,0.150000,0.150000}%
\pgfsetstrokecolor{textcolor}%
\pgfsetfillcolor{textcolor}%
\pgftext[x=0.282889in,y=0.952701in,,bottom,rotate=90.000000]{\color{textcolor}\rmfamily\fontsize{9.600000}{11.520000}\selectfont soiling loss [\%]}%
\end{pgfscope}%
\begin{pgfscope}%
\pgfpathrectangle{\pgfqpoint{0.682116in}{0.549401in}}{\pgfqpoint{2.673884in}{0.806599in}}%
\pgfusepath{clip}%
\pgfsetroundcap%
\pgfsetroundjoin%
\pgfsetlinewidth{1.204500pt}%
\definecolor{currentstroke}{rgb}{0.003922,0.450980,0.698039}%
\pgfsetstrokecolor{currentstroke}%
\pgfsetdash{}{0pt}%
\pgfpathmoveto{\pgfqpoint{0.803656in}{1.319336in}}%
\pgfpathlineto{\pgfqpoint{1.109840in}{1.319336in}}%
\pgfpathlineto{\pgfqpoint{1.186386in}{1.114687in}}%
\pgfpathlineto{\pgfqpoint{1.240529in}{1.256245in}}%
\pgfpathlineto{\pgfqpoint{1.485103in}{1.319336in}}%
\pgfpathlineto{\pgfqpoint{1.513107in}{1.087274in}}%
\pgfpathlineto{\pgfqpoint{1.632594in}{1.086509in}}%
\pgfpathlineto{\pgfqpoint{1.802489in}{0.586065in}}%
\pgfpathlineto{\pgfqpoint{1.819291in}{1.177680in}}%
\pgfpathlineto{\pgfqpoint{1.867833in}{1.134057in}}%
\pgfpathlineto{\pgfqpoint{1.931310in}{1.319336in}}%
\pgfpathlineto{\pgfqpoint{2.289770in}{1.319336in}}%
\pgfpathlineto{\pgfqpoint{2.310306in}{0.978554in}}%
\pgfpathlineto{\pgfqpoint{2.368183in}{0.904979in}}%
\pgfpathlineto{\pgfqpoint{2.384985in}{1.125218in}}%
\pgfpathlineto{\pgfqpoint{2.435394in}{0.985572in}}%
\pgfpathlineto{\pgfqpoint{2.452197in}{1.319336in}}%
\pgfpathlineto{\pgfqpoint{2.498871in}{1.130981in}}%
\pgfpathlineto{\pgfqpoint{2.732243in}{1.319336in}}%
\pgfpathlineto{\pgfqpoint{2.933877in}{1.319336in}}%
\pgfpathlineto{\pgfqpoint{2.999221in}{0.938949in}}%
\pgfpathlineto{\pgfqpoint{3.139244in}{0.896744in}}%
\pgfpathlineto{\pgfqpoint{3.174717in}{1.319336in}}%
\pgfpathlineto{\pgfqpoint{3.234460in}{1.319336in}}%
\pgfpathlineto{\pgfqpoint{3.234460in}{1.319336in}}%
\pgfusepath{stroke}%
\end{pgfscope}%
\begin{pgfscope}%
\pgfsetrectcap%
\pgfsetmiterjoin%
\pgfsetlinewidth{1.003750pt}%
\definecolor{currentstroke}{rgb}{1.000000,1.000000,1.000000}%
\pgfsetstrokecolor{currentstroke}%
\pgfsetdash{}{0pt}%
\pgfpathmoveto{\pgfqpoint{0.682116in}{0.549401in}}%
\pgfpathlineto{\pgfqpoint{0.682116in}{1.356000in}}%
\pgfusepath{stroke}%
\end{pgfscope}%
\begin{pgfscope}%
\pgfsetrectcap%
\pgfsetmiterjoin%
\pgfsetlinewidth{1.003750pt}%
\definecolor{currentstroke}{rgb}{1.000000,1.000000,1.000000}%
\pgfsetstrokecolor{currentstroke}%
\pgfsetdash{}{0pt}%
\pgfpathmoveto{\pgfqpoint{3.356000in}{0.549401in}}%
\pgfpathlineto{\pgfqpoint{3.356000in}{1.356000in}}%
\pgfusepath{stroke}%
\end{pgfscope}%
\begin{pgfscope}%
\pgfsetrectcap%
\pgfsetmiterjoin%
\pgfsetlinewidth{1.003750pt}%
\definecolor{currentstroke}{rgb}{1.000000,1.000000,1.000000}%
\pgfsetstrokecolor{currentstroke}%
\pgfsetdash{}{0pt}%
\pgfpathmoveto{\pgfqpoint{0.682116in}{0.549401in}}%
\pgfpathlineto{\pgfqpoint{3.356000in}{0.549401in}}%
\pgfusepath{stroke}%
\end{pgfscope}%
\begin{pgfscope}%
\pgfsetrectcap%
\pgfsetmiterjoin%
\pgfsetlinewidth{1.003750pt}%
\definecolor{currentstroke}{rgb}{1.000000,1.000000,1.000000}%
\pgfsetstrokecolor{currentstroke}%
\pgfsetdash{}{0pt}%
\pgfpathmoveto{\pgfqpoint{0.682116in}{1.356000in}}%
\pgfpathlineto{\pgfqpoint{3.356000in}{1.356000in}}%
\pgfusepath{stroke}%
\end{pgfscope}%
\end{pgfpicture}%
\makeatother%
\endgroup%

%% file: Meyers_Soiling.bbl
\begin{thebibliography}{10}
\providecommand{\url}[1]{#1}
\csname url@samestyle\endcsname
\providecommand{\newblock}{\relax}
\providecommand{\bibinfo}[2]{#2}
\providecommand{\BIBentrySTDinterwordspacing}{\spaceskip=0pt\relax}
\providecommand{\BIBentryALTinterwordstretchfactor}{4}
\providecommand{\BIBentryALTinterwordspacing}{\spaceskip=\fontdimen2\font plus
\BIBentryALTinterwordstretchfactor\fontdimen3\font minus
  \fontdimen4\font\relax}
\providecommand{\BIBforeignlanguage}[2]{{%
\expandafter\ifx\csname l@#1\endcsname\relax
\typeout{** WARNING: IEEEtran.bst: No hyphenation pattern has been}%
\typeout{** loaded for the language `#1'. Using the pattern for}%
\typeout{** the default language instead.}%
\else
\language=\csname l@#1\endcsname
\fi
#2}}
\providecommand{\BIBdecl}{\relax}
\BIBdecl

\bibitem{Ilse2019}
\BIBentryALTinterwordspacing
K.~Ilse, L.~Micheli, B.~W. Figgis, K.~Lange, D.~Daßler, H.~Hanifi,
  F.~Wolfertstetter, V.~Naumann, C.~Hagendorf, R.~Gottschalg, and J.~Bagdahn,
  ``Techno-economic assessment of soiling losses and mitigation strategies for
  solar power generation,'' \emph{Joule}, vol.~3, no.~10, pp. 2303--2321, 2019.
  [Online]. Available:
  \url{https://www.sciencedirect.com/science/article/pii/S2542435119304222}
\BIBentrySTDinterwordspacing

\bibitem{SEIA2021}
M.~Davis, C.~Smith, B.~White, R.~Goldstein, X.~Sun, M.~Cox, G.~Curtin,
  R.~Manghani, S.~Rumery, C.~Silver, and J.~Baca, \emph{{U.S. {S}olar market
  insight executive summary, 2020 year in review}}.\hskip 1em plus 0.5em minus
  0.4em\relax Wood Mackenzie and SEIA, 2021.

\bibitem{Meyers2022}
\BIBentryALTinterwordspacing
B.~Meyers and S.~Boyd, ``Signal decomposition using masked proximal
  operators,'' pp. 1--60, feb 2022. [Online]. Available:
  \url{http://arxiv.org/abs/2202.09338}
\BIBentrySTDinterwordspacing

\bibitem{Meyers2020b}
B.~{Meyers}, E.~{Apostolaki-Iosifidou}, and L.~{Schelhas}, ``Solar data tools:
  Automatic solar data processing pipeline,'' in \emph{2020 47th IEEE
  Photovoltaic Specialists Conference (PVSC)}, 2020, pp. 0655--0656.

\bibitem{solar-data-tools}
\BIBentryALTinterwordspacing
B.~Meyers, ``solar-data-tools,'' may 2022. [Online]. Available:
  \url{http://dx.doi.org/10.5281/zenodo.6450368}
\BIBentrySTDinterwordspacing

\bibitem{deceglie2018}
M.~Deceglie, L.~Micheli, and M.~Muller, ``Quantifying soiling loss directly
  from pv yield,'' \emph{IEEE Journal of Photovoltaics}, vol.~8, no.~2, pp.
  547--551, 2018.

\bibitem{skomedal2019}
{\AA}.~Skomedal, H.~Haug, and E.~S. Marstein, ``Endogenous soiling rate
  determination and detection of cleaning events in utility-scale pv plants,''
  \emph{IEEE Journal of Photovoltaics}, vol.~9, no.~3, pp. 858--863, 2019.

\bibitem{Skomedal2020}
\BIBentryALTinterwordspacing
A.~Skomedal and M.~Deceglie, ``{Combined Estimation of Degradation and Soiling
  Losses in Photovoltaic Systems},'' \emph{IEEE Journal of Photovoltaics},
  vol.~10, no.~6, pp. 1788--1796, nov 2020. [Online]. Available:
  \url{https://ieeexplore.ieee.org/document/9186286/}
\BIBentrySTDinterwordspacing

\bibitem{Micheli2021}
L.~Micheli, M.~Theristis, A.~Livera, J.~Stein, G.~Georghiou, M.~Muller,
  F.~Almonacid, and E.~Fernandez, ``Improved {PV} soiling extraction through
  the detection of cleanings and change points,'' \emph{IEEE Journal of
  Photovoltaics}, vol.~11, no.~2, pp. 519--526, 2021.

\bibitem{Hastie2009}
\BIBentryALTinterwordspacing
T.~Hastie, R.~Tibshirani, and J.~Friedman, \emph{The Elements of Statistical
  Learning}, ser. Springer Series in Statistics.\hskip 1em plus 0.5em minus
  0.4em\relax New York, NY: Springer New York, dec 2009. [Online]. Available:
  \url{http://ieeexplore.ieee.org/document/6727256/
  http://link.springer.com/10.1007/978-0-387-84858-7}
\BIBentrySTDinterwordspacing

\bibitem{Koenker1978}
\BIBentryALTinterwordspacing
R.~Koenker and G.~Bassett, ``Regression quantiles,'' \emph{Econometrica},
  vol.~46, no.~1, p.~33, jan 1978. [Online]. Available:
  \url{https://www.jstor.org/stable/1913643
  https://www.jstor.org/stable/1913643?origin=crossref}
\BIBentrySTDinterwordspacing

\bibitem{Koenker2001}
\BIBentryALTinterwordspacing
R.~Koenker and K.~F. Hallock, ``Quantile regression,'' \emph{Journal of
  Economic Perspectives}, vol.~15, no.~4, pp. 143--156, nov 2001. [Online].
  Available: \url{https://pubs.aeaweb.org/doi/10.1257/jep.15.4.143}
\BIBentrySTDinterwordspacing

\bibitem{Boyd2018}
S.~Boyd and L.~Vandenberghe, \emph{{Introduction to Applied Linear Algebra}},
  2018.

\bibitem{convex_opt}
------, \emph{Convex optimization}.\hskip 1em plus 0.5em minus 0.4em\relax
  Cambridge University Press, 2009.

\bibitem{diamond2016cvxpy}
S.~Diamond and S.~Boyd, ``{CVXPY}: {A} {P}ython-embedded modeling language for
  convex optimization,'' \emph{Journal of Machine Learning Research}, vol.~17,
  no.~83, pp. 1--5, 2016.

\bibitem{agrawal2018rewriting}
A.~Agrawal, R.~Verschueren, S.~Diamond, and S.~Boyd, ``A rewriting system for
  convex optimization problems,'' \emph{Journal of Control and Decision},
  vol.~5, no.~1, pp. 42--60, 2018.

\bibitem{Andersen2000}
\BIBentryALTinterwordspacing
E.~D. Andersen and K.~D. Andersen, ``{The Mosek Interior Point Optimizer for
  Linear Programming: An Implementation of the Homogeneous Algorithm},'' in
  \emph{High performance optimization}, 2000, pp. 197--232. [Online].
  Available: \url{http://link.springer.com/10.1007/978-1-4757-3216-0{\_}8}
\BIBentrySTDinterwordspacing

\bibitem{convexopt}
S.~Boyd and L.~Vandenberghe, ``{Convex Optimization},'' in \emph{Cambridge
  University Press}, 2004.

\bibitem{numpy-diff}
\BIBentryALTinterwordspacing
``Numpy.diff,'' \emph{{NumPy} v1.22 Manual}. [Online]. Available:
  \url{https://numpy.org/doc/stable/reference/generated/numpy.diff.html}
\BIBentrySTDinterwordspacing

\bibitem{Kam-Lum2021}
\BIBentryALTinterwordspacing
E.~Kam-Lum, B.~E. Meyers, D.~Cosme, B.~Aissa, and G.~Scabbia, ``{Soiling Rate
  Determination from Referenced Systems in Desert Climate using PVInsight
  Soiling Algorithm},'' in \emph{2021 IEEE 48th Photovoltaic Specialists
  Conference (PVSC)}, no. July 2019.\hskip 1em plus 0.5em minus 0.4em\relax
  IEEE, jun 2021, pp. 2552--2554. [Online]. Available:
  \url{https://ieeexplore.ieee.org/document/9518459/}
\BIBentrySTDinterwordspacing

\bibitem{pvsc-shade}
B.~Meyers and D.~F. Florez~Rodriguez, ``Estimation of shade losses in unlabeled
  {PV} data,'' \emph{Submitted to PVSC49}, June 2022.

\bibitem{pvsc-shade-david}
D.~F. Florez~Rodriguez and B.~Meyers, ``Solar panel power simulation for shade
  detection,'' \emph{Submitted to PVSC49}, June 2022.

\bibitem{soiling-notebook}
\BIBentryALTinterwordspacing
B.~Meyers, ``Soiling analysis demonstration,'' \emph{solar-data-tool GitHub
  repository}, May 2022. [Online]. Available:
  \url{https://github.com/slacgismo/solar-data-tools/blob/shade-dev/notebooks/Soiling\_analysis\_demonstration.ipynb}
\BIBentrySTDinterwordspacing

\bibitem{Waskom2021}
\BIBentryALTinterwordspacing
M.~L. Waskom, ``seaborn: statistical data visualization,'' \emph{Journal of
  Open Source Software}, vol.~6, no.~60, p. 3021, 2021. [Online]. Available:
  \url{https://doi.org/10.21105/joss.03021}
\BIBentrySTDinterwordspacing

\end{thebibliography}
